# Vector-Based Approach to the Stoichiometric Analysis of Multicomponent Chemical Reactions: The Case of Black Powder

Pavlo Kozub[1], Nataliia Yilmaz[2], Svitlana Kozub[3]

1. Kharkiv National University of Radio Electronics, Kharkiv, Ukraine
2. School of Computer and Communication Sciences, École Polytechnique Fédérale de Lausanne (EPFL), CH-1015 Lausanne, Switzerland
3. Medical and bioorganic chemistry department, Kharkiv National Medical University, Kharkiv, Ukraine

**Corresponding author:** pkozub@pkozub.com

**Abstract**

The study demonstrates the capabilities of a vector-based approach for calculating stoichiometric coefficients in chemical equations, using black powder as an illustrative example. A method is proposed for selecting and constraining intermediate interactions between reactants, as well as for identifying final products. It is shown that even a small number of components can lead to a large number of final and intermediate products. Through concrete calculations, a correlation is established between the number of possible chemical equations and the number of reactants. A methodology is proposed for computing all possible chemical equations within a reaction system for arbitrary component ratios, enabling the derivation of all feasible chemical reactions. Additionally, a method is developed for calculating the chemical composition for a fixed set of reactants, allowing for the evaluation of the set of products resulting from all possible chemical interactions given a specified initial composition.

**Introduction**

Balancing chemical reactions is one of the fundamental components of modern chemistry, without which it is impossible to perform any technological calculations related to chemical processes. Although this problem has been studied for over 150 years, the currently available methods for determining stoichiometric coefficients still do not meet the requirements of scientific and technical practice. Contemporary methods are either too simplistic to solve complex equations or too complex to be implemented in practice [1, 2].

A key distinction of the vector-based approach to calculating the material balance of complex chemical processes is its ability to compute the complete set of linearly independent reactions between reactants and products, regardless of the number of components and without imposing constraints on the presence of reactants or products with identical compositions [3, 4, 5].

The capabilities of this approach can be demonstrated for various reaction systems, but a particularly illustrative example is the reaction system formed during the combustion of black powder, which consists of only three components: approximately 75% potassium nitrate, 15% charcoal, and 10% sulfur [6].

This composition is considered traditional; however, there are numerous variants of black powder that differ in the proportions of the components (sometimes by as much as 5%) as well as in the presence of additional substances and additives [7, 8].

Depending on the type of carbonaceous material used, different forms of powder are obtained: black powder (pure carbon), brown powder (from brown coal), or chocolate powder (from coal tar). The addition of a hydrogen-containing component increases the energetic output of the powder, making it suitable for artillery applications.

In theory, sulfur can be entirely replaced with organic substances – for example, potassium nitrate and sugar (75% potassium nitrate, 25% sugar). Carbon can also be fully substituted with another carbon-containing compound, such as potassium carbonate (55% nitrate, 18% sulfur, and 27% potassium carbonate). To further increase combustion energy, aluminum is sometimes added to black powder.

In any case, the initial mixture must contain an oxidizing agent (potassium nitrate) and a reducing agent (sulfur, carbon, or organic substances).

Thus, the actual reactive mixture for black powder may contain not merely three components, but significantly more:

$KNO_3$ – an oxidizing agent, serves as a source of oxygen. It is sometimes used in combination with other nitrates – $NH_4NO_3$, $NaNO_3$, $Ca(NO_3)_2$, $Ba(NO_3)_2$ – which influence the burning rate.

C – a reducing agent, responsible for the formation of explosive gases. In the presence of sufficient oxygen, it is oxidized to $CO_2$; in oxygen-deficient conditions, to CO.

S – a reducing agent, necessary for binding potassium in the form of $K_2S$, or in excess oxygen conditions, in the form of $K_2SO_4$.

$H_2O$ – may appear during storage (due to hydration) or be intentionally added to reduce the burning rate; it may also be present as part of other nitrates.

$C_nH_mO$ – organic substances such as resins, paraffins, waxes, sugar, and wood powder may be added both to modify the combustion process and to alter physical properties (e.g., reduce caking, hygroscopicity, facilitate granulation, etc.).

Al – used primarily to increase combustion temperature.

In the traditional composition of black powder, it is assumed that the main combustion products are only three substances: $CO_2$ (carbon dioxide), $N_2$ (nitrogen), and $K_2S$ (potassium sulfide).

However, it is logically reasonable to assume that, in reality, the addition of hydrogen-containing compounds – such as organic substances, ammonium nitrate, or nitrate hydrates – results in the formation of water as a product of chemical reactions. Incomplete combustion of carbon compounds leads to the formation of CO, and in the case of insufficient potassium ions, carbonates, sulfates, and sulfites may also form.

This is supported by experimental investigations of the combustion products of black powder, yet such outcomes are still scarcely represented in the form of a comprehensive list of possible chemical interactions, as their calculation using traditional methods is nearly infeasible.

**Identification of Reactants and Interactions in the Reaction System**

The principal advantage of the vector-based approach lies in its ability to generate **all possible linearly independent chemical equations** (i.e., combinations of reactants with a constant number of atoms in both the initial and final states) for a system with an arbitrary number of reactants.

The number of such equations increases approximately in geometric progression relative to the number of participating reactants. Although this number can be theoretically estimated using mathematical methods, in practice, it is significantly lower due to constraints of a chemical nature.

For example, actual chemical reactions are usually limited in the number of participating reactants; it is extremely rare for both oxidizing and reducing agents to appear simultaneously among the products; and reactants and products should not possess identical compositions (though they may be isomers or substances in different physical states).

Therefore, the most effective method for investigating a complex reaction system is the gradual increase of its complexity.

In the case of black powder, a reasonable starting point is the traditional composition of the initial reactants, which ensures the formation of only the principal combustion products.

$$KNO_3,C,S \Rightarrow K_2S,CO_2,N_2.$$

Such a composition can be achieved only at a single ratio of the initial reactants.

$$2KNO_3+3C+S=3CO_2+N_2+K_2S$$

A critically important parameter for the practical application of black powder is the volume of gases produced during the chemical interaction, which, for this particular composition of the starting material, amounts to $\Delta V = 89.6\ \mathrm{L \cdot mol^{-1}}$ of reaction mixture or $\Delta V = 332\ \mathrm{cm^3 \cdot g^{-1}}$ of the initial components.

In cases of poor mixing of the components (e.g., coarse granularity, inadequate homogenization of the reactants), reactions involving only two components may occur. Such interactions are also possible when the ratio of the initial reactants deviates from stoichiometric values, meaning that the actual volume of combustion gases may differ from the calculated values. This variation is determined by incomplete combustion reactions, as well as by interactions between intermediate species and the initial reactants (reactants) or final products.

In the case of potassium nitrate ($KNO_3$) deficiency, incomplete oxidation of carbon may occur, leading to the formation of CO.

If sulfur (S) is deficient, potassium carbonate ($K_2CO_3$) may form.

In the case of carbon (C) deficiency, potassium sulfate ($K_2SO_4$) and potassium sulfite ($K_2SO_3$) may form.

If both carbon and sulfur are deficient, sodium nitrite ($NaNO_2$) and molecular oxygen ($O_2$) may be produced.

$$KNO_3,C,S \Rightarrow K_2S,CO_2,N_2+CO,K_2CO_3,K_2SO_3,K_2SO_4,K_2NO_2,O_2$$

Thus, as a result of black powder combustion, at least 15 stoichiometrically feasible chemical reactions are possible. However, some of these can be excluded due to the actual incompatibility of certain reaction products (e.g., $O_2$ and $K_2SO_3$, S and $O_2$, CO and $O_2$, etc.), which ultimately leaves only 12 reactions.

$$2KNO_3+S=N_2+K_2SO_4+O_2$$
$$2KNO_3+C=N_2+K_2CO_3+1.5O_2$$
$$2KNO_3+2.5C=1.5CO_2+N_2+K_2CO_3$$
$$2KNO_3+C+S=CO_2+N_2+K_2SO_4$$
$$2KNO_3+1.5C+S=1.5CO_2+N_2+K_2SO_3$$
$$4KNO_3=N_2+2K_2NO_2+4O_2$$
$$2KNO_3+4C=N_2+3CO+K_2CO_3$$
$$2KNO_3+3C+S=3CO_2+N_2+K_2S$$
$$4KNO_3+4C=4CO_2+N_2+2K_2NO_2$$
$$2KNO_3+3C+S=N_2+3CO+K_2SO_3$$
$$4KNO_3+8C=N_2+8CO+2K_2NO_2$$
$$2KNO_3+6C+S=N_2+K_2S+6CO$$

At elevated temperatures, intermediate products interact both with each other and with the initial reactants. At temperatures above 400 °C, potassium nitrite ($KNO_2$) is formed. Above 440 °C, potassium oxide ($K_2O$) is produced. Above 500 °C, potassium sulfide reacts with oxygen to form sulfate ($K_2SO_4$). Above 900 °C, the sulfate reacts with carbon to yield sulfide and carbon monoxide (CO). Above 600 °C, sulfite undergoes disproportionation into sulfide and sulfate. Above 300 °C, sulfur reacts with oxygen to form sulfur dioxide ($SO_2$). Above 400 °C, sulfur dioxide reacts with oxygen to form sulfur trioxide ($SO_3$). Also above 400 °C, sulfur dioxide reacts with carbon to produce elemental sulfur and carbon dioxide ($CO_2$).

For such a complex reaction system, a total of 198 chemical reactions are possible, but this set can be reduced by retaining only 5 substances as final products.

$$KNO_3,C,S+CO_2,K_2S,CO,K_2CO_3,K_2SO_3,K_2SO_4,KNO_2,K_2O,SO_2,SO_3 \Rightarrow K_2S,CO_2,N_2+K_2CO_3,K_2SO_4$$

The addition of organic substances leads to their decomposition with the formation of hydrogen, water, and carbon. At 250 °C, organic compounds decompose with the release of $H_2$. At 300 °C, aromatic compounds begin to form, having a general empirical formula of $(CH)_n$. At 400 °C, elemental carbon (C) starts to form. At the same temperature (400 °C), hydrogen and organic compounds react with oxygen to form $H_2O$.

When water is present in the mixture – either as a result of organic compound decomposition or due to ambient moisture – nitrogen oxides, sodium hydroxide, and sulfuric acid are produced.

Above 100 °C, sodium nitrite decomposes in the presence of water to yield $NO_2$, NO, and $K_2O$. The resulting $K_2O$ immediately reacts with $SO_2$, $SO_3$, and $CO_2$ to form $K_2SO_3$, $K_2SO_4$, and $K_2CO_3$, respectively. $K_2CO_3$, in the presence of sulfur oxides, further reacts to form $K_2SO_3$ and $K_2SO_4$. In turn, $K_2SO_3$, in the presence of oxygen or $SO_3$, converts into $K_2SO_4$. The addition of aluminum or iron leads to the formation of their respective oxides.

Thus, the actual combustion process of black powder is significantly more complex than a single chemical reaction, and requires a specialized computational approach and methodology.

**Calculation of Reactions for Arbitrary Component Compositions**

The traditional formulation of black powder consists of three reactants and is designed to yield three primary products.

$$KNO_3,C,S=>K_2S,CO_2,N_2$$

For such a composition of the reaction mixture, only one reaction can exist.

$$2KNO_3+3C+S=K_2S+3CO_2+N_2$$

However, if additional possible products are included – those that arise from deviations from this composition and are terminal (final) under maximum temperature conditions – the system expands by at least four more substances.

$$KNO_3+C+S=>K_2S,CO_2,N_2+K_2O,\ O_2,\ K_2SO_4,\ K_2CO_3$$

As a result, five additional reactions are obtained, which correspond to mixtures with an excess of certain components relative to the traditional composition.

$$2KNO_3=N_2+K_2O+2.5O_2$$
$$2KNO_3+S=N_2+O_2+K_2SO_4$$
$$2KNO_3+C=N_2+1.5O_2+K_2CO_3$$
$$2KNO_3+2.5C=1.5CO_2+N_2+K_2CO_3$$
$$2KNO_3+C+S=CO_2+N_2+K_2SO_4$$

It should be noted that this is an exclusive list of chemical equations involving these components. Since potassium oxide cannot exist freely in an environment containing other acid-forming components, it can be excluded from the set of final products—along with oxygen, which can exist only as a result of the decomposition of pure potassium nitrate. As a result, only three reactions remain that lead to complete consumption of the initial reactants. All other compositional ratios will involve intermediate products.

This does not mean that such reactions are impossible or do not occur in practice, but the amount of intermediate products should decrease over time, in contrast to terminal products.

When paraffins, fats, or waxes (represented in the reaction as $CH_2$) are added to the reaction mixture, an additional terminal product, $H_2O$, appears.

$$KNO_3+C+S+CH_2=>K_2S,CO_2,N_2+K_2SO_4,\ K_2CO_3+H_2O$$

As a result, additional reactions are obtained:

$$3KNO_3+1.5S+CH_2=CO_2+1.5N_2+1.5K_2SO_4+H_2O$$

$$2KNO_3+S+2CH_2=K_2S+2CO_2+N_2+2H_2O$$
$$3KNO_3+2.5CH_2=CO_2+1.5N_2+1.5K_2CO_3+2.5H_2O$$
$$4KNO_3+C+S+3CH_2=K_2S+3CO_2+2N_2+K_2CO_3+3H_2O$$
$$4.4KNO_3+C+3CH_2=1.8CO_2+2.2N_2+2.2K_2CO_3+3H_2O$$
$$4.5KNO_3+1.25S+2.5CH_2=1.5CO_2+2.25N_2+1.25K_2SO_4+K_2CO_3+2.5H_2O$$
$$5KNO_3+2.5S+3CH_2=K_2S+3CO_2+2.5N_2+1.5K_2SO_4+3H_2O$$
$$5KNO_3+S+4.5CH_2=K_2S+3CO_2+2.5N_2+1.5K_2CO_3+4.5H_2O$$
$$6KNO_3+2S+3CH_2=2CO_2+3N_2+2K_2SO_4+K_2CO_3+3H_2O$$
$$6.8KNO_3+S+6CH_2=K_2S+3.6CO_2+3.4N_2+2.4K_2CO_3+6H_2O$$
$$8KNO_3+C+3S+3CH_2=3CO_2+4N_2+3K_2SO_4+K_2CO_3+3H_2O$$
$$14KNO_3+7S+6CH_2=K_2S+6CO_2+7N_2+6K_2SO_4+6H_2O$$

When carbohydrates (with the general formula $CH_2O$) are added to the reaction mixture, a similar set of possible reactions is obtained, but with a different composition of reaction products.

$$KNO_3+C+S+CH_2O=>K_2S,CO_2,N_2+K_2SO_4,\ K_2CO_3+H_2O$$

$$2KNO_3+2.5CH_2O=1.5CO_2+N_2+K_2CO_3+2.5H_2O$$
$$2KNO_3+S+CH_2O=CO_2+N_2+K_2SO_4+H_2O$$
$$2KNO_3+S+3CH_2O=K_2S+3CO_2+N_2+3H_2O$$
$$3.2KNO_3+C+3CH_2O=2.4CO_2+1.6N_2+1.6K_2CO_3+3H_2O$$
$$4.4KNO_3+S+6CH_2O=K_2S+4.8CO_2+2.2N_2+1.2K_2CO_3+6H_2O$$
$$5KNO_3+1.5S+4CH_2O=3CO_2+2.5N_2+1.5K_2SO_4+K_2CO_3+4H_2O$$
$$5KNO_3+C+1.5S+3CH_2O=3CO_2+2.5N_2+1.5K_2SO_4+K_2CO_3+3H_2O$$
$$6KNO_3+2C+S+6CH_2O=K_2S+6CO_2+3N_2+2K_2CO_3+6H_2O$$
$$6KNO_3+S+6CH_2O=4CO_2+3N_2+K_2SO_4+2K_2CO_3+6H_2O$$
$$6KNO_3+S+8CH_2O=K_2S+6CO_2+3N_2+2K_2CO_3+8H_2O$$
$$7KNO_3+2.5S+5CH_2O=4CO_2+3.5N_2+2.5K_2SO_4+K_2CO_3+5H_2O$$
$$8KNO_3+4S+6CH_2O=K_2S+6CO_2+4N_2+3K_2SO_4+6H_2O$$
$$14KNO_3+S+18CH_2O=K_2S+12CO_2+7N_2+6K_2CO_3+18H_2O$$

To increase the energetic output of black powder, aluminum is sometimes added, resulting in the formation of only one new product in the reaction mixture: $Al_2O_3$.

$$KNO_3+C+S+Al=>K_2S,CO_2,N_2+K_2SO_4,\ K_2CO_3+H_2O+Al_2O_3$$

Since it, like sulfur and carbon, acts as a reducing agent, the number of new reactions increases by only three.

$$2KNO_3+C+2Al=N_2+K_2CO_3+Al_2O_3$$
$$3KNO_3+1.5S+2Al=1.5N_2+1.5K_2SO_4+Al_2O_3$$
$$2KNO_3+S+4Al=K_2S+N_2+2Al_2O_3$$

Thus, although deviations from the traditional composition of black powder may lead to the formation of a large number of intermediate products in the final reaction mixture, the introduction of constraints that allow only stable products can significantly reduce the number of possible chemical interactions between reactants.

To compute equations involving intermediate products, it is sufficient to include them in the list of reactants and products; however, this results in the number of possible reactions increasing to several hundred.

**Calculation of Reactions for a Fixed Component Composition**

The presented calculations may be of interest for theoretical studies; however, for practical applications, it is more useful to compute the composition of the reaction mixture for a given ratio of components in the initial mixture, or to determine the required ratios of initial components to achieve a desired composition of the reaction mixture.

A fundamental principle of the vector-based approach is that, for a fixed ratio between reactants, only a finite number of reactions exist that satisfy the material balance between the initial reactants and the products of chemical interaction. This number is exhaustive and may, in some cases, be zero. However, if the initial reactants are included among the products of the interaction, then there will always exist at least one set of product ratios that satisfies the material balance of the chemical transformation.

For this purpose, it is advisable to include the initial reactants as part of the final products, and to represent black powder as a composite substance with a fixed composition.

$$[(KNO_3)_2C_3S](KNO_3)_nC_mS_k=K_2S,CO_2,N_2+K_2SO_4,\ K_2CO_3+H_2O+KNO_3,C,S$$

where $[(KNO_3)_2C_3S]$ – the classical composition of black powder;
*n, m, k* – deviations in the quantities of individual components.

This approach makes it possible to investigate the effect of excess amounts of each reactant on the composition of the combustion products of the reactant mixture. In turn, this enables the calculation of the volumes of gases produced by each reaction, normalized to the mass of the initial reaction mixture ($cm^3/g$).

For example, with an excess of 0.5 mol of $KNO_3$ – ($[(KNO_3)_2C_3S](KNO_3)_{0.5}$) – the following reactant ratios can be obtained:

$4[(KNO_3)_2C_3S_1](KNO_3)_{0.5}=2.75K_2S+11CO_2+5N_2+1.25K_2SO_4+K_2CO_3$ ($\Delta V$ = 280 $cm^3 \cdot g^{-1}$)
$12[(KNO_3)_2C_3S_1](KNO_3)_{0.5}=22CO_2+15N_2+K_2SO_4+14K_2CO_3+11S$ ($\Delta V$ = 215 $cm^3 \cdot g^{-1}$)
$4[(KNO_3)_2C_3S_1](KNO_3)_{0.5}=5.5CO_2+5N_2+4K_2SO_4+K_2CO_3+5.5C$ ($\Delta V$ = 183 $cm^3 \cdot g^{-1}$)

For an excess of 1 mol of $KNO_3$, the following reactant ratios can be obtained.

$2.67[(KNO_3)_2C_3S_1](KNO_3)_1=K_2S+6.67CO_2+4N_2+1.67K_2SO_4+1.33K_2CO_3$ ($\Delta V$ = 241 $cm^3 \cdot g^{-1}$)
$2[(KNO_3)_2C_3S_1](KNO_3)_1=4CO_2+3N_2+K_2SO_4+2K_2CO_3+S$ ($\varDelta$ $\Delta V$ = 211 $cm^3 \cdot g^{-1}$)
$2[(KNO_3)_2C_3S_1](KNO_3)_1=3.5CO_2+3N_2+2K_2SO_4+K_2CO_3+1.5C$ ($\Delta V$ = 196 $cm^3 \cdot g^{-1}$)
$2.5[(KNO_3)_2C_3S_1](KNO_3)_2=5.5CO_2+4.5N_2+2.5K_2SO_4+2K_2CO_3+KNO_3$ ($\Delta V$ = 190 $cm^3 \cdot g^{-1}$)

$1.25[(KNO_3)_2C_3S_1](KNO_3)_4=2.75CO_2+2.25N_2+1.25K_2SO_4+K_2CO_3+3KNO_3$ ($\Delta V$ = 133 $cm^3 \cdot g^{-1}$)

Similar calculations can be performed for excess amounts of sulfur.

With a sulfur excess of 0.5 mol relative to the traditional composition, it is possible to achieve its complete utilization.

$1.33[(KNO_3)_2C_3S_1]S_{0.5}=K_2S+4CO_2+2N_2+K_2SO_4$ ($\Delta V$ = 353 $cm^3 \cdot g^{-1}$)
$2[(KNO_3)_2C_3S_1]S_{0.5}=4CO_2+3N_2+K_2SO_4+2K_2CO_3+2S$ ($\Delta V$ = 282 $cm^3 \cdot g^{-1}$)
$[(KNO_3)_2C_3S_1]S_{0.5}=1.5CO_2+1.5N_2+1.5K_2SO_4+1.5C$ ($\Delta V$ = 188 $cm^3 \cdot g^{-1}$)

However, when the excess is increased to 1 mol, there are no stoichiometrically feasible reactions that allow for complete utilization of sulfur.

$2[(KNO_3)_2C_3S_1]S_1=1.5K_2S+6CO_2+3N_2+1.5K_2SO_4+S$ ($\Delta V$ = 334 $cm^3 \cdot g^{-1}$)
$1.33[(KNO_3)_2C_3S_1]S_2=K_2S+4CO_2+2N_2+K_2SO_4+2S$ ($\Delta V$ = 317 $cm^3 \cdot g^{-1}$)
$2[(KNO_3)_2C_3S_1]S_1=4CO_2+3N_2+K_2SO_4+2K_2CO_3+3S$ ($\Delta V$ = 260 $cm^3 \cdot g^{-1}$)
$2[(KNO_3)_2C_3S_1]S_1=3CO_2+3N_2+3K_2SO_4+3C+S$ ($\Delta V$ = 223 $cm^3 \cdot g^{-1}$)
$2[(KNO_3)_2C_3S_1]S_2=4CO_2+3N_2+K_2SO_4+2K_2CO_3+5S$ ($\Delta V$ = 235 $cm^3 \cdot g^{-1}$)
$[(KNO_3)_2C_3S_1]S_2=1.5CO_2+1.5N_2+1.5K_2SO_4+1.5C+1.5S$ ($\Delta V$ = 201 $cm^3 \cdot g^{-1}$)

When excess carbon is added, it also remains unreacted or leads to the presence of unconsumed sulfur.

$2.67[(KNO_3)_2C_3S_1]C_{0.5}=1.67K_2S+8CO_2+4N_2+K_2SO_4+1.33K_2CO_3$ ($\Delta V$ = 365 $cm^3 \cdot g^{-1}$)
$6[(KNO_3)_2C_3S_1]C_{0.5}=13CO_2+9N_2+K_2SO_4+8K_2CO_3+5S$ ($\Delta V$ = 298 $cm^3 \cdot g^{-1}$)
$2[(KNO_3)_2C_3S_1]C_{0.5}=3.5CO_2+3N_2+2K_2SO_4+K_2CO_3+2.5C$ ($\Delta V$ = 264 $cm^3 \cdot g^{-1}$)

An increase in excess leads to an increase in the number of possible chemical reactions.

$8[(KNO_3)_2C_3S_1]C_1=7K_2S+28CO_2+12N_2+K_2SO_4+4K_2CO_3$ ($\Delta V$ = 397 $cm^3 \cdot g^{-1}$)
$12[(KNO_3)_2C_3S_1]C_1=10K_2S+41CO_2+18N_2+2K_2SO_4+6K_2CO_3+C$ ($\Delta V$ = 391 $cm^3 \cdot g^{-1}$)
$2[(KNO_3)_2C_3S_1]C_1=K_2S+6CO_2+3N_2+2K_2CO_3+S$ ($\Delta V$ = 357 $cm^3 \cdot g^{-1}$)
$4[(KNO_3)_2C_3S_1]C_1=K_2S+9CO_2+6N_2+3K_2SO_4+2K_2CO_3+5C$ ($\Delta V$ = 298 $cm^3 \cdot g^{-1}$)
$4[(KNO_3)_2C_3S_1]C_1=9CO_2+6N_2+6K_2CO_3+C+4S$ ($\Delta V$ = 298 $cm^3 \cdot g^{-1}$)
$2[(KNO_3)_2C_3S_1]C_1=3.5CO_2+3N_2+2K_2SO_4+K_2CO_3+3.5C$ ($\Delta V$ = 258 $cm^3 \cdot g^{-1}$)

However, further increase in the excess once again reduces their number.

$2[(KNO_3)_2C_3S_1]C_2=2K_2S+7.5CO_2+3N_2+K_2CO_3+1.5C$ ($\Delta V$ = 417 $cm^3 \cdot g^{-1}$)
$[(KNO_3)_2C_3S_1]C_2=2.25CO_2+1.5N_2+1.5K_2CO_3+1.25C+S$ ($\Delta V$ = 286 $cm^3 \cdot g^{-1}$)
$2[(KNO_3)_2C_3S_1](KNO_3)_0C_2=3.5CO_2+3N_2+2K_2SO_4+K_2CO_3+5.5C$ ($\Delta V$ = 248 $cm^3 \cdot g^{-1}$)

When carbohydrates (e.g., sugar) are added to the traditional composition, it is observed that their excess leads to incomplete consumption of the initial reactants. However, the unreacted component may be carbon, sulfur, or both reactants simultaneously.

$[(KNO_3)_2C_3S_1](CH_2O)=K_2S+3CO_2+N_2+H_2O+C$ ($\Delta V$ = 394 $cm^3 \cdot g^{-1}$)
$4[(KNO_3)_2C_3S_1](CH_2O)=3K_2S+10.5CO_2+4N_2+K_2CO_3+4H_2O+4.5C+S$ ($\Delta V$ = 365 $cm^3 \cdot g^{-1}$)
$3[(KNO_3)_2C_3S_1](CH_2O)=2K_2S+7CO_2+3N_2+K_2SO_4+3H_2O+5C$ ($\Delta V$ = 342 $cm^3 \cdot g^{-1}$)
$7[(KNO_3)_2C_3S_1](CH_2O)=K_2S+12CO_2+7N_2+6K_2CO_3+7H_2O+10C+6S$ ($\Delta V$ = 293 $cm^3 \cdot g^{-1}$)
$[(KNO_3)_2C_3S_1](CH_2O)=1.5CO_2+N_2+K_2CO_3+H_2O+1.5C+S$ ($\Delta V$ = 276 $cm^3 \cdot g^{-1}$)
$3[(KNO_3)_2C_3S_1](CH_2O)=4CO_2+3N_2+K_2SO_4+2K_2CO_3+3H_2O+6C+2S$ ($\Delta V$ = 263 $cm^3 \cdot g^{-1}$)
$7[(KNO_3)_2C_3S_1](CH_2O)=K_2S+9CO_2+7N_2+6K_2SO_4+7H_2O+19C$ ($\Delta V$ = 259 $cm^3 \cdot g^{-1}$)
$4[(KNO_3)_2C_3S_1](CH_2O)=4.5CO_2+4N_2+3K_2SO_4+K_2CO_3+4H_2O+10.5C+S$ ($\Delta V$ = 246 $cm^3 \cdot g^{-1}$)
$[(KNO_3)_2C_3S_1](CH_2O)=CO_2+N_2+K_2SO_4+H_2O+3C$ ($\Delta V$ = 237 $cm^3 \cdot g^{-1}$)

The analysis of the presented reactions reveals several important points. First, each possible composition of the initial reactants results in multiple chemical reactions. Second, these reactions produce different specific volumes of gaseous products (ΔV). Third, for most reactions, this volume is lower than that of the traditional composition, although there are also reactions that lead to a higher amount of gaseous products.

The analysis further shows that the presence of potassium sulfide among the reaction products increases the gas volume, while the presence of sulfur results in a higher specific gas yield than carbon.

This, in turn, implies that the final composition of the reaction products cannot be predicted solely based on the initial composition of the reaction mixture, as the products are formed simultaneously through different reactions, and their relative proportions are determined by factors such as temperature, pressure, and time.

**Calculation of Reactions for Fractional Component Compositions**

For practical calculations, it is more convenient to use an initial composition with fractional coefficients. This allows for the computation of all possible reactions and all possible combinations of products for a given initial composition, under the assumption that only stable reaction products are formed.

The number of reactions can be quickly determined based on the weight composition of the powder and the molar masses of the reactants.

$$x = \frac{2.7 \cdot w_{KNO3}}{101}, \qquad y = \frac{2.7 \cdot w_C}{12}, \qquad z = \frac{2.7 \cdot w_S}{32}$$

where

*w(X)* is the mass fraction of substance *X* in the mixture ($KNO_3$, C, S), expressed as a percentage (%).

The coefficients 101, 12, and 32 correspond to the molar masses of the substances ($KNO_3$, C, S) (in g/mol).

For the classical formulation — 75% $KNO_3$, 15% C, 10% S — we obtain $[(KNO_3)_2C_3S_1]$.

And according to the transformation scheme

$$[(KNO_3)_2C_3S_1]=>K_2S,CO_2,N_2+K_2SO_4, K_2CO_3+H_2O+KNO3,C,S$$

only a single possible chemical equation is obtained.

$$[(KNO_3)_2C_3S_1] = K_2S+3CO_2+N_2$$

For the traditional composition of 75% $KNO_3$, 15% C, and 10% S, we obtain $[(KNO_3)_2C_{3.4}S_{0.8}]$, which does not precisely correspond to the theoretical stoichiometric ratio of the components.

[(KNO3)2C3.4S0.8]=>K2S,CO2,N2+K2SO4, K2CO3+H2O+KNO3,C,S

As a result, we now obtain three possible chemical equations with additional products.

5[(KNO3)2C3.4S0.8]=4K2S+13.5CO2+5N2+K2CO3+2.5C
1.25[(KNO3)2C3.4S0.8]=1.88CO2+1.25N2+1.25K2CO3+1.13C+S
5[(KNO3)2C3.4S0.8]=5.5CO2+5N2+4K2SO4+K2CO3+10.5C

With an excess of sulfur in the powder composition — 75% $KNO_3$, 10% C, 15% S — the resulting mixture has the composition $[(KNO_3)_2C_{2.2}S_{1.3}]$, which deviates even further from the classical formulation.

[(KNO3)2C2.2S1.3] =>K2S,CO2,N2+K2SO4, K2CO3+H2O+KNO3,C,S

and also results in the presence of three chemical equations.

3.33[(KNO3) 2C2.2S1.3]=2K2S+7.33CO2+3.33N2+1.33K2SO4+S
3.33[(KNO3) 2C2.2S1.3]=3.33CO2+3.33N2+3.33K2SO4+4C+S
5[(KNO3) 2C2.2S1.3]=7CO2+5N2+K2SO4+4K2CO3+5.5S

The addition of hydrocarbons (paraffin, oil, fat) introduces hydrogen atoms into the system. For the composition 75% $KNO_3$, 10% C, 15% S, and 4% paraffin, the resulting chemical composition is ($[(KNO_3)_2C_{2.3}S_1(CH_2)_{0.8}]$

[(KNO3)2C2.3S1(CH2)0.8] =>K2S,CO2,N2+K2SO4, K2CO3+H2O+KNO3,C,S

which indicates the possibility of a significantly larger number of reactions.

1.25[(KNO3)2C2.3S1(CH2)0.8]=1.38CO2+1.25N2+1.25K2CO3+H2O+1.25C+1.25S
1.67[(KNO3)2C2.3S1(CH2)0.8]=CO2+1.67N2+1.67K2SO4+1.33H2O+4.17C
2[(KNO3)2C2.3S1(CH2)0.8]=2K2S+5.2CO2+2N2+1.6H2O+C
2.5[(KNO3)2C2.3S1(CH2)0.8]=1.5K2S+4.5CO2+2.5N2+K2SO4+2H2O+3.25C
2.5[(KNO3)2C2.3S1(CH2)0.8]=2.25CO2+2.5N2+K2SO4+1.5K2CO3+2H2O+4C+1.5S
2.5(KNO3)2C2.3S1(CH2)0.8]=N2+K2SO4+2H2O+3KNO3+7.75C+1.5S
7.5[(KNO3)2C2.3S1(CH2)0.8]=K2S+6.5CO2+7.5N2+6.5K2SO4+6H2O+16.75C

3.75[(KNO3)2C2.3S1(CH2)0.8]=2.75CO2+3.75N2+2.75K2SO4+K2CO3+3H2O+7.87C+S
3.75[(KNO3)2C2.3S1(CH2)0.8]=2.75K2S+8.25CO2+3.75N2+K2CO3+3H2O+2.37C+S
7.5[(KNO3)2C2.3S1(CH2)0.8]=K2S+9.75CO2+7.5N2+6.5K2CO3+6H2O+7C+6.5S

The addition of sugar or cellulose instead of paraffin (75% $KNO_3$, 10% C, 15% S, 4% sugar) yields the chemical composition $[(KNO_3)_2C_{2.3}S_1(C_6H_{12}O_6)_{0.1}]$.

[(KNO3)2C2.3S1(C6H12O6)0.1] =>K2S,CO2,N2+K2SO4, K2CO3+H2O+KNO3,C,S

which yields a similar set of chemical equations, but with slightly different coefficients.

1.54[(KNO3)2C2.3S1(C6H12O6)0.1]=1.5CO2+N2+K2CO3+9.23H2O+11.35C+1.54S
2.86[(KNO3)2C2.3S1(C6H12O6)0.1]=1.86CO2+1.86N2+1.86K2SO4+17.14H2O+23.86C+S
2.86[(KNO3)2C2.3S1(C6H12O6)0.1]=1.86K2S+5.57CO2+1.86N2+17.14H2O+20.14C+S
3.91[(KNO3)2C2.3S1(C6H12O6)0.1]=1.54K2S+6.12CO2+2.54N2+K2CO3+23.43H2O+28.03C+2.37S
3.91[(KNO3)2C2.3S1(C6H12O6)0.1]=1.54K2S+5.62CO2+2.54N2+K2SO4+23.43H2O+29.53C+1.37S
6.27[(KNO3)2C2.3S1(C6H12O6)0.1]=K2S+7.62CO2+4.08N2+3.08K2CO3+37.63H2O+45.76C+5.27S
6.27[(KNO3)2C2.3S1(C6H12O6)0.1]=5.62CO2+4.08N2+K2SO4+3.08K2CO3+37.63H2O+47.76C+5.27S
8.64[(KNO3)2C2.3S1(C6H12O6)0.1]=6.12CO2+5.62N2+4.62K2SO4+K2CO3+51.83H2O+70.64C+4.02S
8.64[(KNO3)2C2.3S1(C6H12O6)0.1]=K2S+7.62CO2+5.62N2+4.62K2SO4+51.83H2O+70.14C+3.02S

In a similar manner, the composition of the final products can be calculated when part of the potassium nitrate is replaced with ammonium nitrate (50% $KNO_3$, 10% C, 15% S, 25% $NH_4NO_3$). For this component ratio, the composition can be written as $[(KNO_3)_{1.3}C_3S_1(NH_4NO_3)_{0.8}]$, which corresponds to the transformation scheme.

[(KNO3)1.3C3S1(NH4NO3)0.8] =>K2S,CO2,N2+K2SO4, K2CO3+H2O+KNO3,C,S

The formal number of possible reactions in this system is significantly higher (42), but considering the constraints on product composition and the number of reactants, a smaller subset may be used for analyzing the reaction mechanism and for practical calculations.

$1.54[(KNO_3)_{1.3}C_3S_1(NH_4NO_3)_{0.8}]=2.27CO_2+2.54N_2+K_2CO_3+3.08H_2O+1.35C+1.54S$
$2.86[(KNO_3)_{1.3}C_3S_1(NH_4NO_3)_{0.8}]=1.86K_2S+7CO_2+4.71N_2+5.71H_2O+1.57C+S$
$2.86[(KNO_3)_{1.3}C_3S_1(NH_4NO_3)_{0.8}]=3.29CO_2+4.71N_2+1.86K_2SO_4+5.71H_2O+5.29C+S$
$3.3[(KNO_3)_{1.3}C_3S_1(NH_4NO_3)_{0.8}]=4.29CO_2+5.44N_2+1.14K_2SO_4+K_2CO_3+6.6H_2O+4.6C+2.15S$
$3.31[(KNO_3)1.3C3S1*0.8(NH_4NO_3)]=1.15K_2S+6.1CO_2+5.46N_2+K_2SO_4+6.62H_2O+3.82C+1.16S$
$3.31[(KNO_3)_{1.3}C_3S_1(NH_4NO_3)_{0.8}]=4.38CO_2+5.46N_2+K_2SO_4+1.15K_2CO_3+6.62H_2O+4.39C+2.31S$
$3.56[(KNO_3)_{1.3}C_3S_1(NH_4NO_3)_{0.8}]=4.75CO_2+5.87N_2+K_2SO_4+1.31K_2CO_3+7.12H_2O+4.62C+2.56S$
$3.81[(KNO_3)_{1.3}C_3S_1(NH_4NO_3)_{0.8}]=4.88CO_2+6.28N_2+1.48K_2SO_4+K_2CO_3+7.62H_2O+5.54C+2.33S$
$3.81[(KNO_3)_{1.3}C_3S_1(NH_4NO_3)_{0.8}]=1.47K_2S+7.83CO_2+6.28N_2+K_2CO_3+7.62H_2O+2.59C+2.33S$
$4.4[(KNO_3)_{1.3}C_3S_1(NH_4NO_3)_{0.8}]=K_2S+7.06CO_2+7.26N_2+1.86K_2SO_4+8.8H_2O+6.14C+1.54S$
$4.76[(KNO_3)1.3C3S1*0.8(NH_4NO_3)]=2.09K_2S+9.65CO_2+7.85N_2+K_2SO_4+9.51H_2O+4.62C+1.66S$
$5.31[(KNO_3)_{1.3}C_3S_1(NH_4NO_3)_{0.8}]=K_2S+9.33CO_2+8.76N_2+2.45K_2CO_3+10.62H_2O+4.14C+4.31S$
$5.31[(KNO_3)_{1.3}C_3S_1(NH_4NO_3)_{0.8}]=K_2S+8.1CO_2+8.76N_2+2.45K_2SO_4+10.62H_2O+7.82C+1.86S$
$5.66[(KNO_3)_{1.3}C_3S_1(NH_4NO_3)_{0.8}]=2.68K_2S+12.38CO_2+9.35N_2+K_2CO_3+11.33H_2O+3.62C+2.98S$

$5.85[(KNO_3)_{1.3}C_3S_1(NH_4NO_3)_{0.8}]=K_2S+10.12CO_2+9.65N_2+2.8K_2CO_3+11.69H_2O+4.62C+4.85S$
$6.26[(KNO_3)_{1.3}C_3S_1(NH_4NO_3)_{0.8}]=3.07K_2S+13.33CO_2+10.33N_2+K_2SO_4+12.52H_2O+5.44C+2.19S$
$7.46[(KNO_3)_{1.3}C_3S_1(NH_4NO_3)_{0.8}]=1.89K_2S+13.33CO_2+12.31N_2+K_2SO_4+1.96K_2CO_3+14.92H_2O+7.08C+4.57S$
$11.93[(KNO_3)_{1.3}C_3S_1(NH_4NO_3)_{0.8}]=6.75K_2S+27.72CO_2+19.68N_2+K_2CO_3+23.85H_2O+7.06C+5.17S$

Moreover, analysis of this system of equations indicates that all of them contain residual amounts of unreacted carbon and sulfur, meaning that even theoretically, complete combustion of the reactants cannot be achieved with this composition. Therefore, such a formulation is not optimal.

## Conclusions

Thus, using a relatively simple reaction system, the capabilities of the vector-based approach for analyzing chemical interactions have been demonstrated.

It has been shown that even a small number of components can lead to a large number of final and intermediate products.

Through real computational examples, a relationship has been established between the number of possible chemical equations and the number of reactants.

A simple yet effective methodology has been provided for calculating all possible chemical equations within a reaction system for arbitrary component ratios, allowing the enumeration of all feasible chemical reactions.

A method has also been proposed for calculating the chemical composition for a fixed set of reactants, enabling the evaluation of the set of products resulting from possible chemical interactions for a given initial composition.